\documentclass{article}
\usepackage{amsmath}
\usepackage{graphicx}

\title{Strings and Branes under Microscope}
\author{V. Dzhunushaliev
\thanks{E-mail: dzhun@hotmail.kg}}

\date{}

\begin{document}
\maketitle

\begin{center}
\textit{Dept. Phys. and Microel. Engineer., KRSU, Bishkek, \\
Kievskaya Str. 44, 720021, Kyrgyz Republic}
\end{center}

\begin{abstract}
It is shown that in the standard vacuum 5D Kaluza-Klein gravity there is 
wormhole-like solutions which look like strings attached to two 
D-branes. The asymptotical behaviour of the corresponding metric 
is investigated. 
\end{abstract}

\section{Introduction}

In string theory the strings are the matter and consequently the inner 
structure of string is not defined. It is like to the situation in the 
classical and quantum field theories: the point-like electron does not 
have any inner structure. It is well known that the structureless electron 
leads to such difficulties as the infiniteness electron mass, infinities 
in loops of Feynman diagrams and so on. In string theory the pointwise 
electron is dragged out in one dimension and the inner structure remains 
yet unknown. 
\par 
Einstein has offered an idea that pointwise electron has an inner structure 
and it is a bridge (wormhole on the modern language) between remote parts 
of a Universe. Up to now this program is not realized as there are many 
difficulties connected with this idea. For instance the geometrical 
interpretation of the spin is not clear in this electron model. In Ref's 
\cite{dzhun1} it is shown that this idea may be is more applicable for 
strings not for pointlike particles. In these papers it is shown that 
in the standard 5D Kaluza-Klein gravity there is flux tube solutions filled 
with electric and magnetic fields and certain part of this solution can be 
superlong and superthin. The cross section of this flux tube can be choosen 
in the Planck region and consequently the tube can be considered as 
1-dimensional object, namely a $\Delta-$string. $\Delta$ means that the 
ends of the tube\footnote{where the string is attached to an external space} 
are similar to a river delta because a spacetime foam spreads these 
attachment points. 
\par 
In Ref's \cite{dzhun1} the properties of the central part of the solution 
are investigated. In this paper the peripheral parts of the 
$\Delta$-string solution will be studied. We will see that they have 
asymptotically flat spaces (tails) and intermediate regions. By neglecting 
the intermediate regions whole construction is similar to $D-$branes with 
string stretched between them. 

\section{The metric and $\Delta-$string part of solution}

We investigate the 5D metric $G_{AB}, A,B=0,1,2,3,5$ in the following 
form 
\begin{equation}
  ds^2 = \frac{a(r)}{\Delta (r)} dt^2 - dr^2 - a(r) 
  \left(
  d\theta^2 + \sin^2 \theta d\varphi^2 
  \right) - 
  \frac{\Delta (r)}{a(r)} e^{2\psi (r)} 
  \left(
  d\chi + \omega (r) dt + Q \cos \theta d\varphi
  \right)^2 
\label{sec1:10}
\end{equation}
the functions $a(r), \Delta(r)$ and $\psi(r)$ are the even functions and 
consequently the metric has a wormhole-like form; 
$Q$ is the magnetic charge. The form of $G_{5\mu}=(\omega, 0,0, Q\cos \theta )$  $(\mu=0,1,2,3)$ leads to the appearance of the radial electric and magnetic 
F fields. The 5D Einstein's equations are 
\begin{equation}
  R_{AB} - \frac{1}{2} \eta_{AB} R = 0
\label{sec1:20}
\end{equation}
here $A,B$ are 5-bein indices; $R_{AB}$ and $R$ are 5D Ricci tensor and the 
scalar curvature respectively; $\eta_{AB} = diag\{ 1,-1,-1,-1,-1 \}$. 
The 5D Einstein's 
equations for the metric \eqref{sec1:10} are 
\begin{eqnarray}
  R_{15} = \omega'' + \omega'
  \left(
  -\frac{a'}{a} + 2\frac{\Delta'}{\Delta} + 3 \psi '
  \right) &=& 0 ,
\label{sec1:22}\\
  -R_{33} = \frac{a''}{a} +\frac{a'\psi'}{a} -\frac{2}{a} + 
  \frac{Q^2 \Delta e^{2\psi}}{a^3} &=& 0 ,
\label{sec1:24}\\
  R_{11} - R_{55} = \psi'' + {\psi'}^2 + \frac{a' \psi'}{a} - 
  \frac{Q^2 \Delta e^{2\psi}}{2a^3} &=& 0 ,
\label{sec1:26}\\
  2 R_{11} + R_{22} -R_{33} - R_{55} = 
  \frac{\Delta''}{\Delta} - \frac{\Delta' a'}{\Delta a} + 
  3\frac{\Delta' \psi'}{\Delta} + \frac{2}{a} - 6 \frac{a' \psi'}{a} &=& 0 ,
\label{sec1:27}\\
  -R_{55} + R_{22} - 2R_{33} + 2R_{11} = 
  \frac{{\Delta'}^2}{\Delta^2} + \frac{4}{a} - 
  \frac{q^2 e^{-4\psi}}{\Delta^2} - 
  \frac{Q^2 \Delta e^{2\psi}}{a^3} - 6\frac{a' \psi'}{a} - 
  2\frac{\Delta' a'}{\Delta a} + 2\frac{\Delta ' \psi '}{\Delta}&=& 0 .
\label{sec1:28} 
\end{eqnarray} 
The solution of Maxwell equation \eqref{sec1:22} is 
\begin{equation}
  \omega' = \frac{q a e^{-3\psi}}{\Delta^2} 
\label{sec2:60}
\end{equation}
here $q$ is the electric charge. The solutions are parametrized by 
electric $q$ and magnetic $Q$ charges \cite{dzhsin1}: as the relative 
strengths of the electric ans magnetic fields are varied is found that 
the solutions evolve in a following way:
\begin{enumerate} 
\item 
$0 < Q < q$. The solution is a wormhole-like object. The throat between
the surfaces at $\pm r_H$ ($r_H$ is defined as follows: 
$\Delta(\pm r_H) = 0$)  is filled with both electric and magnetic fields. 
The longitudinal distance between the $\pm r_H$ surfaces increases by 
$q \rightarrow Q$.
\item 
$q = Q$. In this case the solution is an infinite flux tube filled
with constant electrical and magnetic fields. The cross-sectional size of 
this solution is constant ($ a= const.$). 
\item 
$0 < q < Q$. In this case we have a singular finite flux tube located 
between two (+) and (-) electrical and magnetic charges located at $\pm r_0$. 
Thus the longitudinal size of this object is finite, but now the cross
sectional size decreases as $r \rightarrow r_0$. 
\end{enumerate} 
By $q=Q$ the solution is the infinite flux tube 
\begin{eqnarray} 
a(r) & = & a(0) = \frac{Q_0^2}{2} = const, 
\label{sec1:30}\\
e^{\psi (r) } & = & \frac{a(0)}{\Delta(r)} = \cosh\frac{r}{\sqrt{a(0)}},
\label{sec1:40}\\
\omega(r) & = & \sqrt{2}\sinh\frac{r}{\sqrt{a(0)}} ,
\label{sec1:45}\\
E &=& \frac{q_0}{a(0)}, \quad 
H = \frac{Q_0}{a(0)}
\label{sec1:50}
\end{eqnarray} 
here we have parallel electric $E$ and magnetic $H$ fields with equal 
$q_0 = Q_0 = \sqrt{2 a(0)}$ electric and magnetic charges, $a_0=a(0)$. 
This solution is 5D analog of the 4D Levi-Civita-Robertson-Bertotti 
solution \cite{levi-civita}-\cite{bertotti}. 
The $\Delta-$string solution is the solution 
with $\delta = 1 - q/q_0 \ll 1$, $q>Q$. In Ref. \cite{dzhun2} this solution 
in the region $|r| < r_H$ is investigated. The topology of this spacetime is 
$R \times S^1 \times S^2 \times [-r_H, +r_H]$, where $R$ is the time 
dimension; $S^1$ is the 5-th dimension; $S^2$ is the 2-sphere spanned 
on $\theta$ and $\varphi$ angles; $r \in [-r_H, +r_H]$ is the radial coordinate. 
The linear sizes of $S^1$ and $S^2$ are in the Planck region. 
\par
In Ref. \cite{dzhun2} the next relation using numerical and approximate 
analytical calculations is derived 
\begin{equation}
  a(r) + \Delta(r) e^{2\psi(r)} \approx 2 a(0) .
\label{sec1:60}
\end{equation}
For $q=0$ and $q=Q$ this relation is exact one but for the $q<Q$ and 
in the region $r \in [-r_H, +r_h]$ this relation was verified using 
numerical and approximate analytical calculations only. This equation 
shows us that in the region $|r| < r_H$ the cross section of the flux 
tube has the same order: $a(0) < a(r) < 2 a(0)$. If 
$a(0) \approx l^2_{Pl}$ then the tube can be considered as one dimensional 
object because $l_{Pl}$ is the least length in the nature. Also the 
investigations show us that the length $L$ of the tube can be arbitrary long 
depending on $\delta$: 
$L \stackrel{\delta \rightarrow 0}{\longrightarrow} \infty$. 
All that allows us to call this flux tube as a string-like object: 
$\Delta-$string. 
\par 
Now we would like to show that the $\Delta-$string solution is nonsingular 
at the points $\pm r_H$ where $\Delta(\pm r_H) = 0$. For this we 
investigate the solution behaviour near to the point $|r| \approx r_H$ where 
\begin{eqnarray}
  a(r) &=& a_0 + a_1 \left ( r-r_H \right ) + 
  a_2 \left ( r-r_H \right )^2 + \cdots ,
    \label{sec1:51}\\
    \psi(r) & = & \psi_H + \psi_1 \left ( r-r_H \right ) + 
    \psi_2 \left ( r-r_H \right )^2 + \cdots ,
    \label{sec1:52}\\
    \Delta(r) & = & \Delta_1 \left( r - r_H \right) + 
    \Delta_1 \Delta_2 \left( r - r_H \right)^2 + \cdots .
    \label{sec1:54}
\end{eqnarray}
The substitution in equations \eqref{sec1:22}-\eqref{sec1:28} gives us the following 
solution
\begin{eqnarray}
    \Delta_1 & = & \pm q e^{-2\psi_H},
    \quad (+) \quad \text{for} \quad r \rightarrow -r_H 
  \quad \text{and} 
  \quad (-) \quad \text{for} \quad r \rightarrow +r_H ,
    \label{sec1:56}\\
    \psi_2 &=& -\psi_1 \frac{a_1 + a_0\psi_1}{2a_0},
    \label{sec1:57}\\   
    \Delta_2 &=& \frac{-3a_0 \psi_1 + a_1}{2 a_0},
    \label{sec1:58}\\   
    a_2 &=& \frac{2 - a_1 \psi_1}{2}.
    \label{sec1:59}
\end{eqnarray}
In this case equation \eqref{sec2:60} has the following behaviour near to the 
points $r=\pm r_H$  
\begin{equation}
    \omega'(r) = \frac{a_0 e^{\psi_H}}{q} \frac{1}{\left( r - r_H \right)^2} + 
    \omega_1 + \mathcal{O}\left( r - r_H \right)
    \label{sec1:59a}
\end{equation}
where $\omega_1$ is some constant depending on $a_{0,1}, \psi_{1,2}, \Delta_{1,2}$. 
It leads to the following $\omega(r)$ behaviour 
\begin{equation}
    \omega(r) = -\frac{a_0 e^{\psi_H}}{q} \frac{1}{\left( r - r_H \right)} + 
    \omega_0 + \mathcal{O}\left( r - r_H \right)
    \label{sec1:59b}
\end{equation}
where $\omega_0$ is some integration constant. The $G_{tt}$ metric component is 
\begin{equation}
    G_{tt} = \frac{a(r)}{\Delta(r)} - 
    \frac{\Delta(r) e^{2 \psi(r)}}{a(r)} \omega^2(r) = 
    -e^{2\psi_H} \frac{2qe^{-\psi_H} \omega_0 - a_1 - a_0 \psi_1}{q} + 
    \mathcal{O} \left( r - r_H \right).
    \label{sec1:59c}
\end{equation}
Then the metric \eqref{sec1:10} has the following approximate behaviour 
near to $r = \pm r_H$ points \begin{equation}
\begin{split}
    &ds^2 = \left[ g_H +    \mathcal{O} \left( r - r_H \right) \right] dt^2 - 
    \mathcal{O} \left( r - r_H \right) 
    \left( d\chi + Q \cos\theta d \varphi \right)^2 - \\
    &\left[ e^{\psi_H} + \mathcal{O} \left( r - r_H \right) \right]
    dt \left( d\chi + Q \cos\theta d \varphi \right) - 
    dr^2 - \left[ a(r_H) + \mathcal{O} \left( r - r_H \right] \right) 
    \left( d\theta^2 + \sin^2 \theta d\varphi^2 \right) \approx \\
    &e^{\psi_H} dt \left( d\chi + Q \cos\theta d \varphi \right) - 
    dr^2 - a(r_H) \left( d\theta^2 + \sin^2 \theta d\varphi^2 \right)
\end{split} 
\label{sec1:59e}
\end{equation}
where 
$g_H = -e^{2\psi_H} \left (2qe^{-\psi_H} \omega_0 - a_1 - a_0 \psi_1 \right )/q$. 
It means that at the points $r = \pm r_H$ the metric \eqref{sec1:10} 
is nonsingular. 

\section{The tails of the $\Delta-$string}

\begin{figure}[h]
  \begin{minipage}[t]{.45\linewidth}
  \begin{center}
    \fbox{
    \includegraphics[height=5cm,width=5cm]{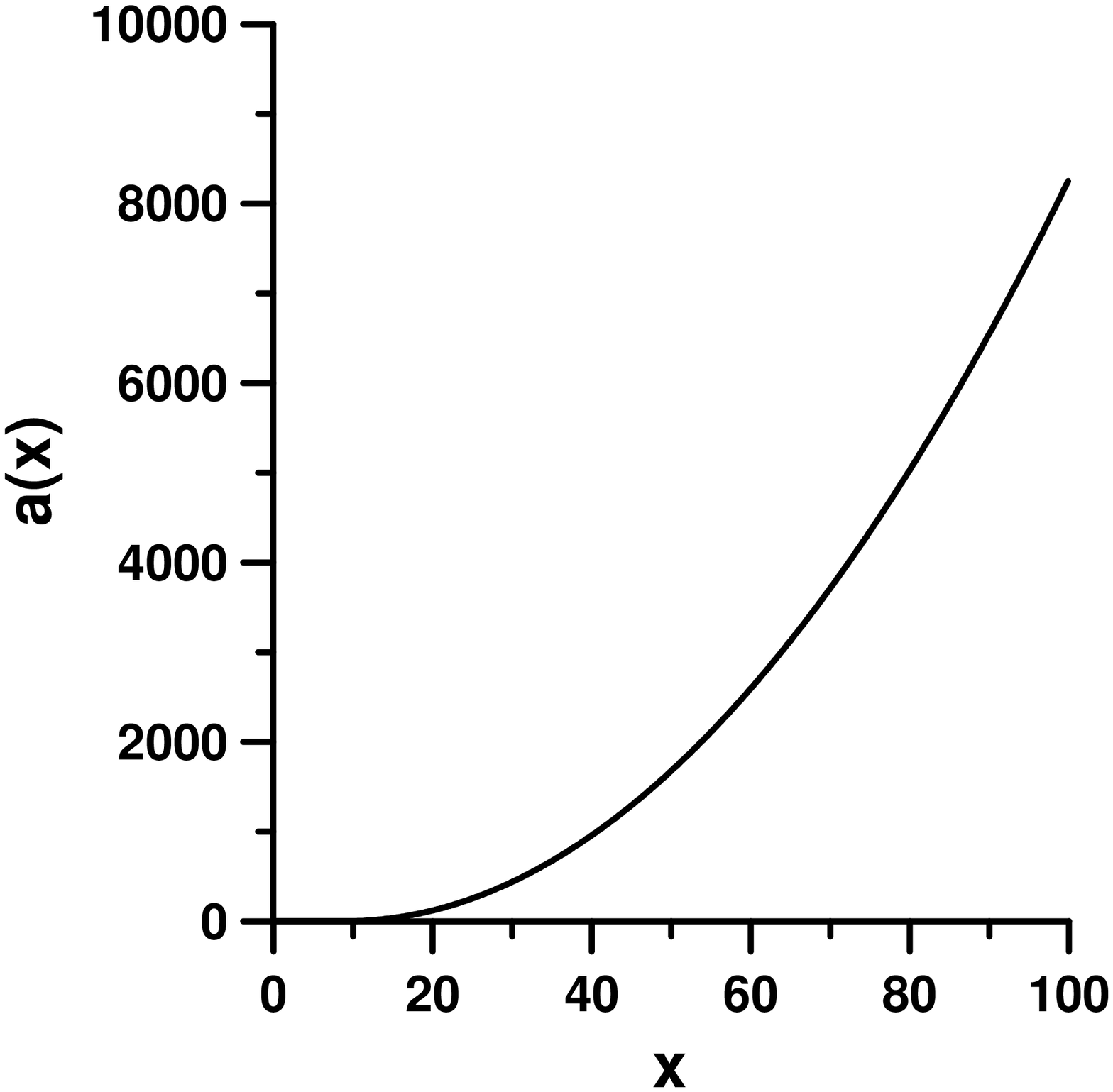}}
    \caption{The function $a(x)$, $\delta \approx 10^{-9}$.
    $x=r/\sqrt{a(0)}$ is the dimensionless radius.}
    \label{fig:a}
  \end{center}  
  \end{minipage}\hfill
  \begin{minipage}[t]{.45\linewidth}
  \begin{center}
    \fbox{
    \includegraphics[height=5cm,width=5cm]{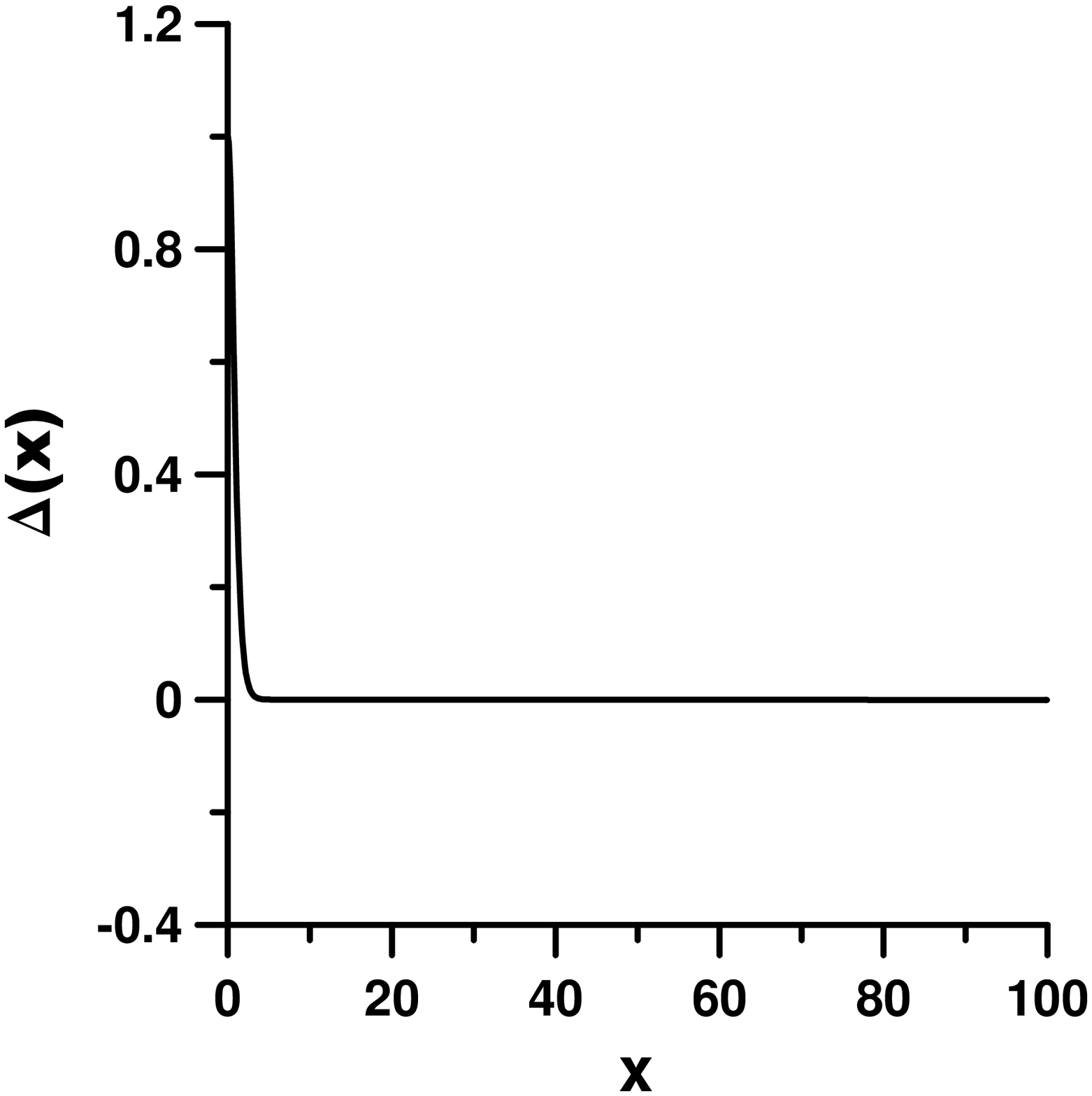}}
    \caption{The function $\Delta(x)$.}
    \label{fig:delta}
  \end{center}  
  \end{minipage} 
\end{figure}

\begin{figure}[h]
  \begin{minipage}[t]{.45\linewidth}
  \begin{center}
    \fbox{
    \includegraphics[height=5cm,width=5cm]{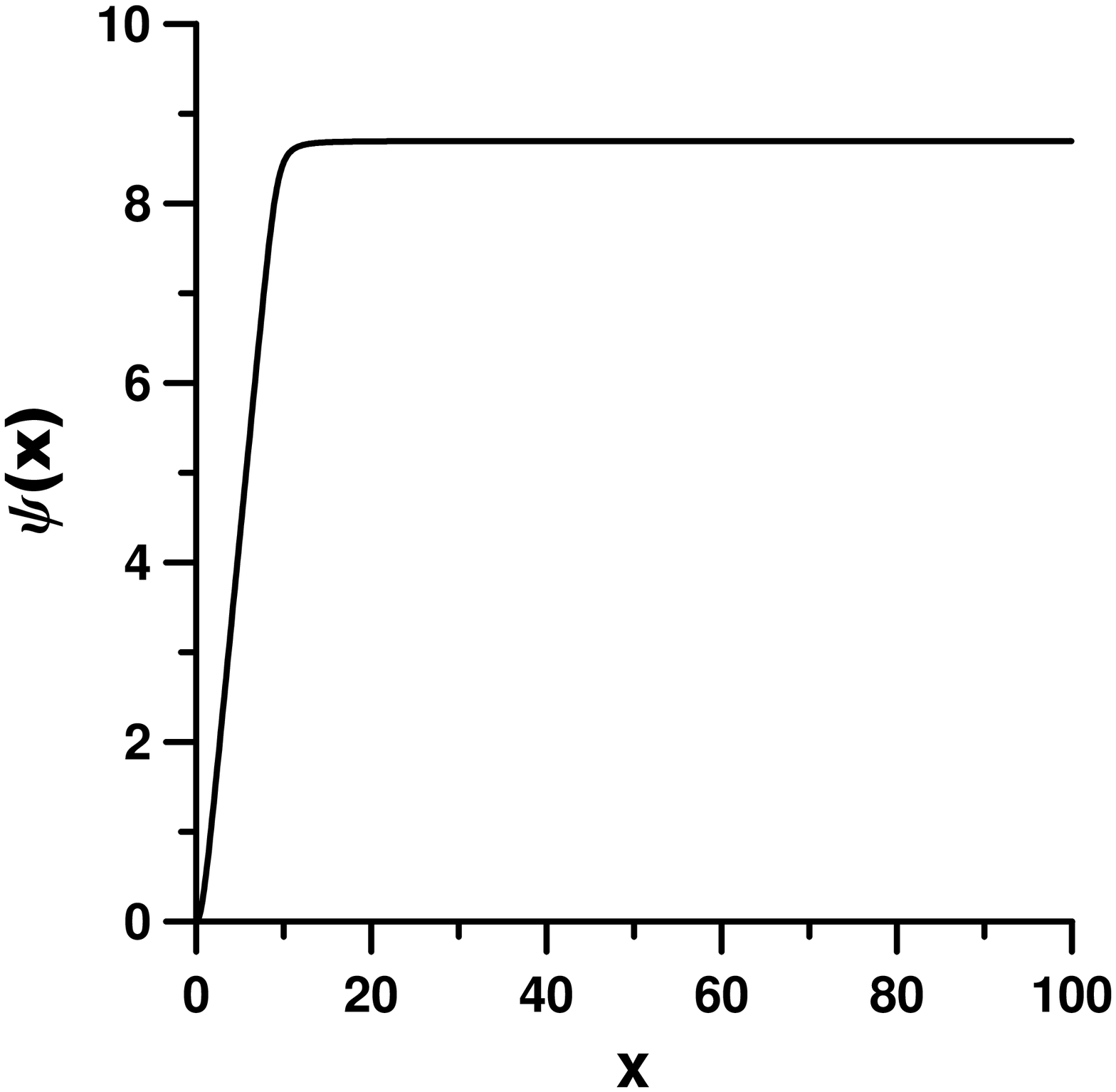}}
    \caption{The function $\psi(x)$.}
    \label{fig:psi}
  \end{center}  
  \end{minipage}\hfill
  \begin{minipage}[t]{.45\linewidth}
  \begin{center}
    \fbox{
    \includegraphics[height=5cm,width=5cm]{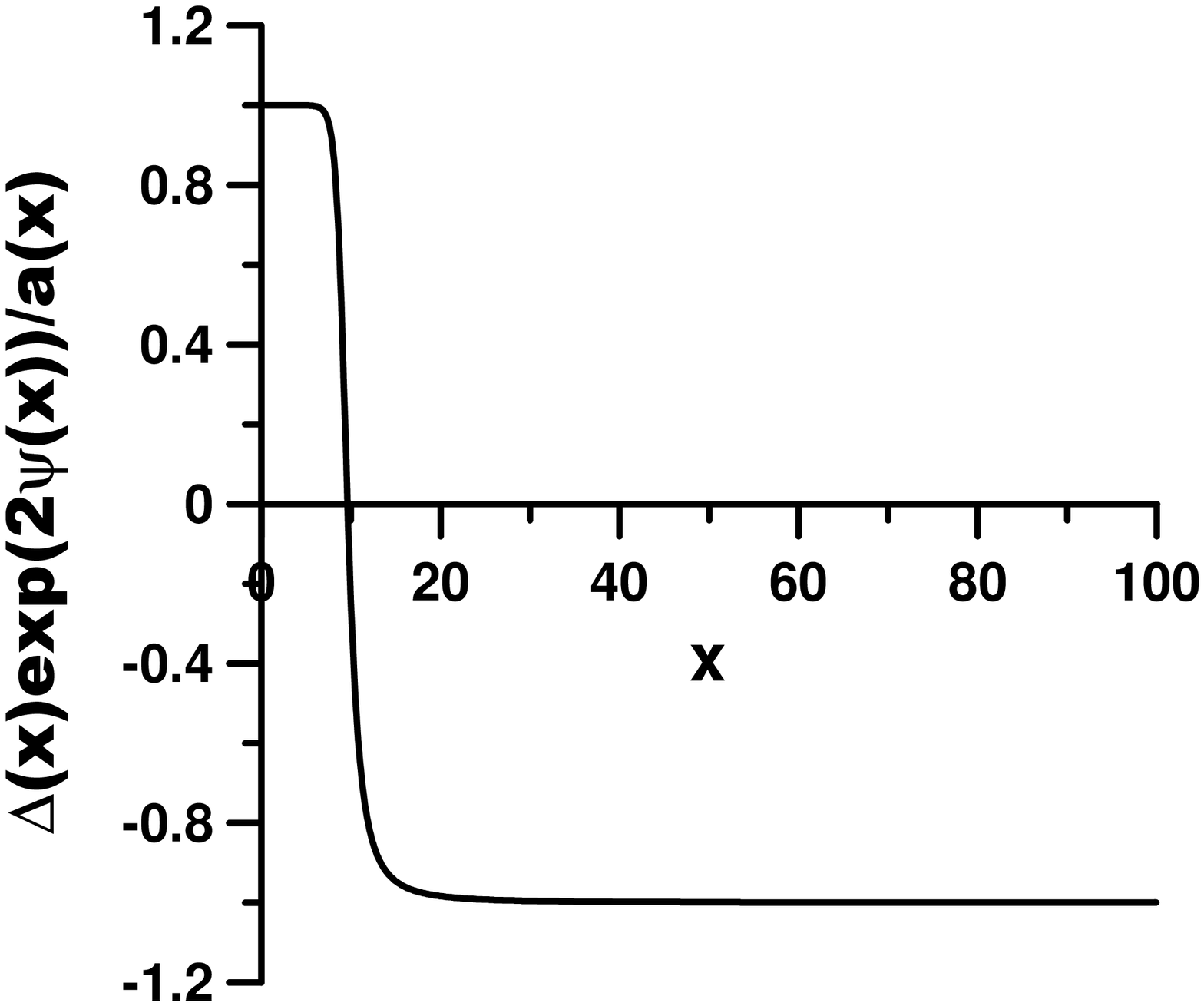}}
    \caption{The function $\frac{\Delta(x) e^{2\psi(x)}}{a(x)}$.}
    \label{fig:prod}
  \end{center}  
  \end{minipage} 
\end{figure}

Now we would like to consider $|r| > r_H$ parts of the solution. The numerical 
investigation of the equations \eqref{sec1:22}-\eqref{sec1:28} are presented 
on Fig's \ref{fig:a}-\ref{fig:prod}, 

We search the asymptotical behavior of the metric in the form 
\begin{eqnarray}
    a(r) & = & r^2 + m_1 r + q_1 + \cdots ,
    \label{sec2:70}\\
    \Delta(r) & = & -\Delta_\infty r^2 + \Delta_\infty m_2 r + 
    \Delta_\infty q_2 + \cdots ,
    \label{sec2:80}\\
    \psi(r) & = & \psi_\infty + \frac{\psi_1}{r^2} + \cdots .
    \label{sec2:90}
\end{eqnarray}
The solution is 
\begin{eqnarray}
    \psi_1 & = & -\frac{Q^2 \Delta_\infty e^{2 \psi_\infty}}{4},
    \label{sec2:100}\\
    q_1 & = & \frac{q^2 e^{-4 \psi_\infty} + 
    3Q^2 \Delta^3_\infty e^{2 \psi_\infty} - 
    \Delta^2_\infty m_2^2 - 2 \Delta^2_\infty m_1 m_2}{4 \Delta^2_\infty}, 
    \label{sec2:110}\\
    q_2 & = &  \frac{q^2 e^{-4 \psi_\infty} - 
    3Q^2 \Delta^3_\infty e^{2 \psi_\infty} - 
    \Delta^2_\infty m_2^2}{4 \Delta^2_\infty}.
    \label{sec2:120}
\end{eqnarray}
The numerical investigation (see, Fig. \ref{fig:prod}) shows us that 
at the infinity 
\begin{equation}
    \frac{\Delta(r) e^{2 \psi(r)}}{a(r)} \approx - \Delta_\infty e^{2\psi_\infty} 
    \left( 1 - \frac{m_1 + m_2}{r} - \frac{m_1^2 - q_1 - q_2 + 2\psi_1}{r^2} \right)
    \rightarrow -1 
    \label{sec2:130}
\end{equation}
and consequently 
\begin{equation}
    \Delta_\infty = e^{-2\psi_\infty} .
    \label{sec2:135}
\end{equation}
After substitution in equations \eqref{sec2:100}-\eqref{sec2:120} we have 
\begin{eqnarray}
    \psi_1 & = & -\frac{Q^2}{4},
    \label{sec2:140}\\
    q_1 & = & \frac{q^2 + 3Q^2 - m_2^2 - 2 m_1 m_2}{4}, 
    \label{sec2:150}\\
    q_2 & = &  \frac{q^2 - 3Q^2 - m_2^2}{4}.
    \label{sec2:160}
\end{eqnarray}
As $q \approx l_{Pl}$ and $Q \approx l_{Pl}$ then on the tails of the 
$\Delta-$string solution ($|r| \gg r_H$) 
\begin{eqnarray}
    \psi_1 & \approx & 0, 
    \label{sec2:170}\\
    q_1 & \approx & -\frac{2 m_1 m_2 + m_2^2}{4},
    \label{sec2:180}\\  
    q_2 & \approx & - \frac{m_2^2}{4}.
    \label{sec2:190}
\end{eqnarray}
This fact allows us to say that the size of the intermediate region 
of the $\Delta-$string solution is $l_{int} \approx m_{1,2}$. 
\par
\begin{figure}[h]
    \begin{center}
     \fbox{
        \includegraphics[width=0.30\textwidth]{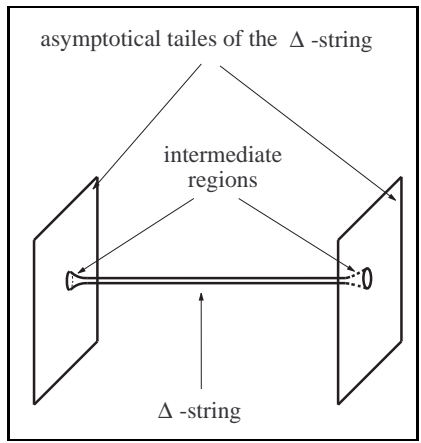}}
    \end{center}
    \caption{The spacetime of the $\Delta-$string.}
    \label{fig:tails}
\end{figure}

 
Thus the $\Delta-$string solution has the superthin and superlong 
flux tube, two intermediate regions and two almost flat spaces, see 
Fig. \ref{fig:tails}. If the cross section of the throat is in the Planck region then 
whole construction looks like a string attached to two D-branes 
(if we neglect of the intermediate regions). Such comparison has the deep physical 
sense as we can isometrically insert the considered spacetime in a sufficiently large Minkowski 
spacetime and then really the throat of the superthin and superlong flux tube is 1D object as Planck 
cell is minimally accessible volume in the nature. 
\par 
Finally we would like to show that the solution considered here is not ordinary black 
metrics presented in review \cite{cveticyoum}. At first we consider the asymptotical 
form of the metric \eqref{sec1:10}
\begin{equation}
\begin{split}
	ds^2 \approx &\left(
		1 - \frac{m_1 + m_2}{r} - \frac{m_1^2 - q_1 - q_2 + 2\psi_1}{r^2}
	\right) 
	\left(
		d \chi + \omega dt + Q \cos \theta d \varphi
	\right)^2 - 
\\
	&\frac{1}{\Delta_\infty}\left(
		1 + \frac{m_1 - m_2}{r} + \frac{q_1 - q_2 + m_2^2}{r^2}
	\right) dt^2 - 
	dr ^2 - r^2 \left( d \theta^2 + \sin^2 \theta d \varphi^2)\right ) = 
\\
	&G_{55} \left(
		d \chi + \omega dt + Q \cos \theta d \varphi
	\right)^2 + g_{\mu \nu} dx^\mu dx^\nu
\end{split}
\label{sec2_200}
\end{equation}
where the asymptotical form of the metric is 
\begin{eqnarray}
	G_{55} &\approx& 1 - \frac{m_1 + m_2}{r} - \frac{m_1^2 - q_1 - q_2 + 2\psi_1}{r^2},
\label{sec2_210}\\
	g_{tt} &\approx& -\frac{1}{\Delta_\infty}\left(
		1 + \frac{m_1 - m_2}{r} + \frac{q_1 - q_2 + m_2^2}{r^2}
	\right),
\label{sec2_220}\\
	g_{\theta \theta} &\approx& -r^2,
\label{sec2_230}\\
	g_{\theta \theta} &\approx& -r^2 \sin^2 \theta .
\label{sec2_240}
\end{eqnarray}
Immediately we see that the $5^{th}$ coordinate $\chi$ in this region becomes 
timelike and the time coordinate $t$ spacelike one. 
\par 
For the clarification of the physical meaning of the asymptotical metric we have to 
bring it to the standard Kaluza-Klein form with new 4D metric $\widetilde{g}_{\mu \nu}$ 
and new electromagnetic potential $\widetilde A_\mu$
\begin{equation}
\begin{split}
	ds^2 \approx &\widetilde{G}_{55} \left(
		d t + \widetilde A_\mu dx^\mu
	\right)^2 + \widetilde{g}_{\mu \nu} dx^\mu dx^\nu = 
\\
	& \left( G_{55} \omega ^2 - g_{tt} \right) 
	\left[
		dt + \frac{G_{55}\omega}{G_{55}\omega^2 - g_{tt}} 
		\left(
			d\chi + Q \cos \theta d \varphi
		\right)
	\right]^2 - 
	\frac{g_{tt} G_{55}}{G_{55}\omega^2 - g_{tt}} 
	\left(
			d\chi + Q \cos \theta d \varphi
		\right)^2 - 
\\
	&dr ^2 - r^2 \left( d \theta^2 + \sin^2 \theta d \varphi^2)\right )	
\end{split}
\label{sec2_250}
\end{equation}
where $\widetilde{G}_{AB}$ ($A,B = 0,1,2,3,5$) is new form of the metric on the tails. 
Here $t$ coordinate is new $5^{th}$  coordinate and $\chi$ is new time coordinate. This form 
of the metric shows that there is a rotation connected with the term $d \chi d \varphi$. 
The $\varphi$-component of electromagnetic potential depends on $r$ and $\theta$ 
coordinates and consequently we have the radial and $\theta$-components of the magnetic field 
in addition to the radial electric field. All of this leads to the appearance of the 
angular momentum density for the electromagnetic field which is the origin for the 
rotation term $d \chi d \varphi$. It allows us to say that 
considered here metric is none of the black hole solutions presented in 
review \cite{cveticyoum}. 

\section{Conclusions}

This investigation shows us that in the pure multidimensional vacuum 
gravity exist wormhole-like solutions which can be approximately considered as two 
D-branes and a string between them with some intermediate regions 
between them. One interesting characteristic 
property of the presented regular solution is that the metric signature is changed 
at some points $\pm r_H$: $(+,-,-,-,-)$ by $|r|<r_H$ is changed on 
$(-,-,-,-,+)$ by $|r| > r_H$. It means that on the $\Delta-$string the 
time coordinate is $t$ coordinate but on the tails (D-branes) the time 
coordinate is 5-th coordinate $\chi$.

\section{Acknowledgment}
I am very grateful to the ISTC grant KR-677 for the financial support.

\end{document}